Report on the ESO and Radionet Workshop on

# Gas and Stars in Galaxies – A Multi-Wavelength 3D Perspective

held at ESO Garching, Germany, 10–13 June 2008

Matt Lehnert[1]
Carlos De Breuck[2]
Harald Kuntschner[2]
Martin Zwaan[2]

[1] Laboratoire d'Etudes des Galaxies, Etoiles, Physique et Instrumentation (GEPI), Observatoire de Paris-Meudon, France
[2] ESO

An overview of the ESO/Radionet workshop devoted to 3D optical/near-infrared and sub-mm/radio observations of gas and stars in galaxies is presented. There will be no published proceedings but presentations are available at *http://www.eso.org/sci/meetings/gal3D2008/program.html*.

The main aim of this ESO/Radionet workshop was to bring together the optical/near-IR and sub-mm/radio communities working on three-dimensional (3D) extragalactic data. The meeting was attended by more than 150 scientists. This article, due to space limitations, provides a, necessarily biased, overview of the meeting. We decided not to publish proceedings, but the presentations are available from the workshop homepage at *http://www.eso.org/sci/meetings/gal3D2008/program.html*. The names of speakers relevant to a topic are included here so that further reference to the presentations can be made through the web page.

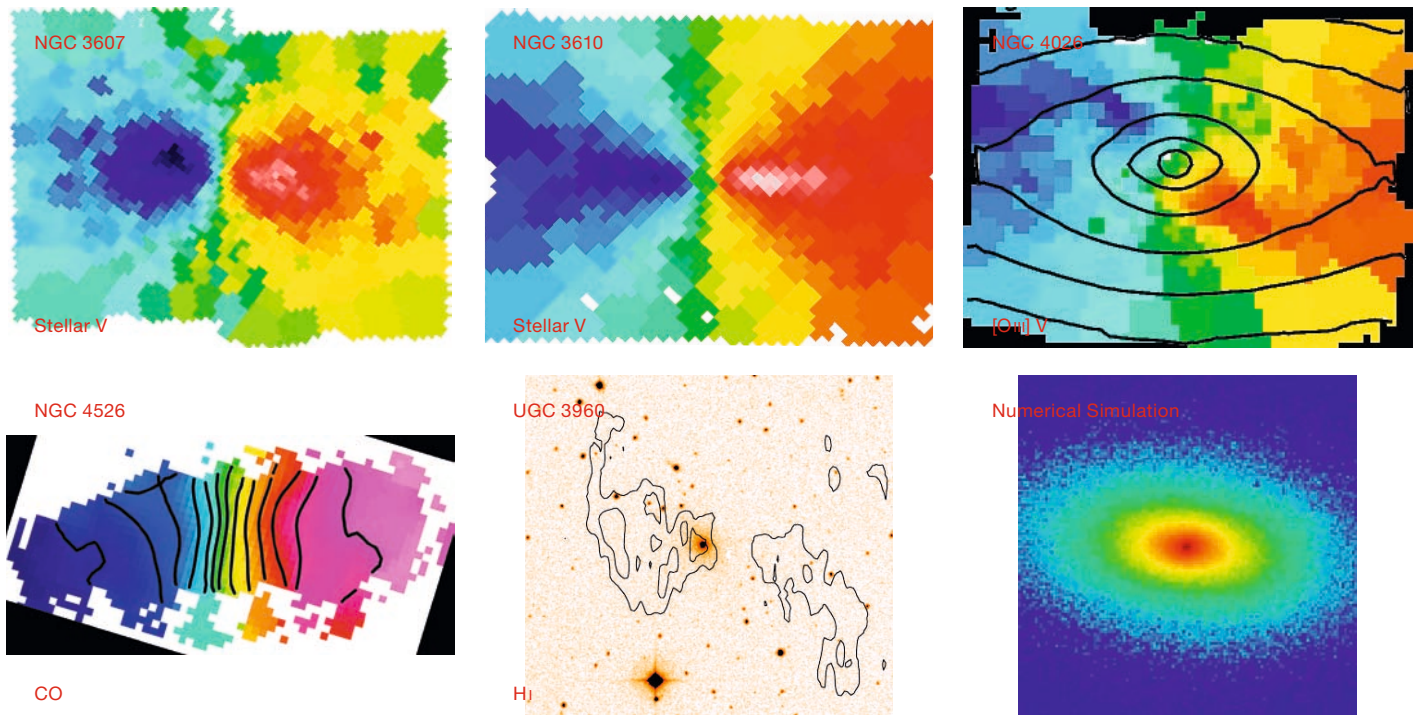

Figure 1. The Atlas$^{3D}$ project *(http://www-astro.physics.ox.ac.uk/atlas3d)* includes a multi-wavelength coverage of a complete sample of nearby early-type galaxies, including optical IFU, CO and H I data combined with a specific effort on numerical simulations. Illustrations of the Atlas$^{3D}$ datasets are shown, from top left to bottom right: stellar velocity maps of NGC 3607 and NGC 3610; ionised gas velocity map of NGC 4026; CO velocity map of NGC 4526; H I contours of UGC 3960; and a projected snapshot of a numerical simulation.

The optical/near-IR community now has access to an increasing number of powerful Integral Field Units (IFUs; see presentation by Eric Emsellem) and the second-generation VLT instruments, as well as the proposed E-ELT instruments, will all have IFU units (Niranjan Thatte). These instruments will thus provide large data cubes sampling the stellar content and the warm/hot ionised gas.

Radio and millimetre interferometers have provided 3D information on gas in galaxies for decades (Thijs van der Hulst). ALMA will – by design – always provide high spatial and spectral resolution data cubes of the cold gas (Robert Laing), allowing the molecular and dust distribution to be traced in galaxies. Future radio facilities (Philip Diamond) will extend current studies of neutral hydrogen out to cosmological distances and will provide information on the cool gas in and around galaxies. All of these devices and techniques are necessary if we are ever going to understand the complex interaction between gas and stars in galaxies.

It was clear from the meeting that we are indeed learning a great deal about the detailed relationships between gas phases in galaxies, how star formation proceeds, and how the global environment within galaxies affects these relationships. Perhaps uniquely emphasised at this meeting was the important role that instrumentation, especially those that provide three-dimensional (spatial and spectral) information, can play in our overall understanding of star formation and galaxy evolution.

### Early- and late-type galaxies

One of the most fascinating themes of the conference was the nature and continuing growth of early type galaxies. The paradigm that early-type galaxies always mean pressure-supported systems with no recent star formation, and certainly no accreted gas, has been consigned to historical novelty. It appears now through the Atlas$^{3D}$ project with the SAURON imaging spectrograph (presentation by Michele Cappellari, see Figure 1) that early-type galaxies show a surprising amount of rotation. This was not apparent previously because of the narrow range of magnitudes and the limited number of the galaxies in earlier surveys. Less luminous, and more numerous early-type galaxies tend to show more rotational support than their more massive and rarer cousins. So most early-type galaxies are lenticulars and not ellipticals. These observations can plausibly be explained by mergers with a range of mass ratios that are typically 1:2 or 1:3 (Thorsten Naab). But of course, the final result depends on the initial orientations and amplitude of the various angular momentum vectors of the progenitors and the orbit (Maxime Bois). However, there were some puzzling, and perhaps alarming, comments that these merger models, while perhaps explaining the large-scale dynamics, cannot account for the orbit families within early-type galaxies. Obviously, the exquisite 3D data that we are able to produce is a real puzzle for modellers (Mathieu Puech).

It also appears that we can watch the growth of structure within early-type galaxies. By studying the CO and H I emis-sion from early-type galaxies, it is possible to observe a relationship between gas content and the dynamical properties and ages of the stars in their circumnuclear regions. For example, even though this relation can only be measured for a relatively short time, due to rapid fading, ~ 20 % of early-type galaxies show evidence for young stars in their circumnuclear regions (Martin Bureau) and overall there appears to be a relationship between young ultraviolet bright discs, CO emission, and young stars and star formation (see Figure 2). In fact, it appears that such star-formation events follow the relation between the star-formation rate and the gas-surface density relation – the so-called Schmidt-Kennicutt relation (Martin Bureau). Even though such observations are just beginning, it is already clear that early-type galaxies can contain a significant amount of neutral hydrogen (10–15 % of which have $10^{9-10}$ solar masses of H I; Raffaella Morganti) with a hint that perhaps galaxies with centrally concentrated H I also have younger stars in their nuclei. These observations may also explain the dichotomy in 'kinematically decoupled cores' (KDCs), that is a core which has different kinematics to the surrounding galaxy. Large KDCs appear to have formed long ago, while smaller KDCs appear to have formed rather recently (results by the SAURON team, reported by Martin Bureau). This difference is likely to be related to recent gas accretion as seen in mm- and cm-wavelength observations.

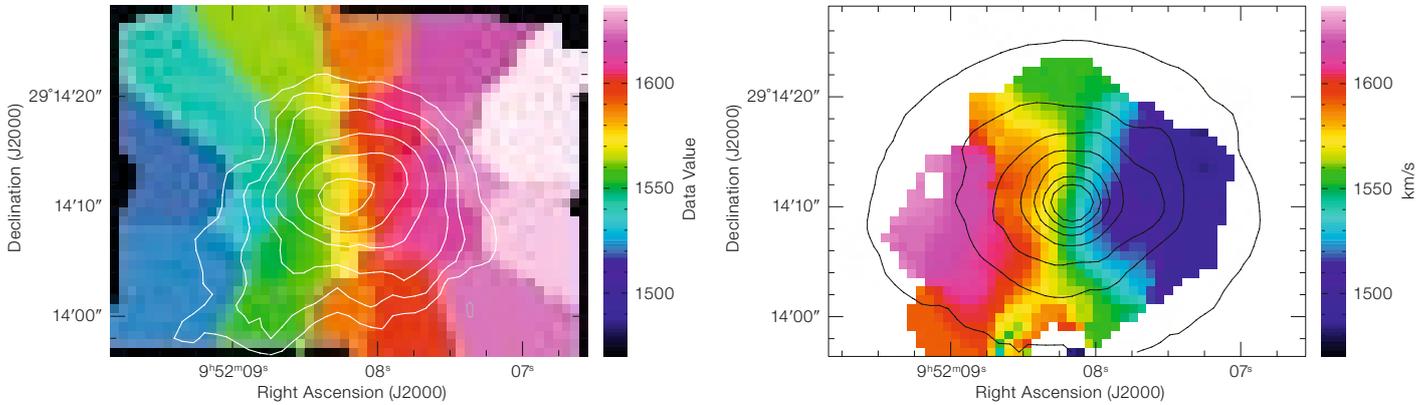

Figure 2. Comparison of CO and optical 3D data for NGC 3032. Left: Stellar mean velocity field, overlaid with contours from the integrated/total CO(1–0) map (Young et al., 2008). Right: CO mean velocity field, overlaid with the optical (roughly V-band) isophotes (Emsellem et al., 2004).

Late-type star-forming galaxies in the local Universe are also full of surprises. There are several theories for explaining how star formation is driven on large scales – from gravitational instabilities, to such instabilities aided by magnetic fields, to large-scale convergent flows or density fluctuations in the gas. Recent multi-wavelength observations in the optical, H I and CO of star-forming galaxies indicate that molecular gas forms with fixed efficiency, that giant molecular cloud populations are universal, but that the formation of molecular clouds clearly depends on the large-scale environment (contribution by Adam Leroy, see Figure 3). This universality of molecular clouds was emphasised by comparing the clouds in our own Milky Way with those of other galaxies (Alberto Bolatto). These data seem to disfavour the primary role of magnetic fields in star formation, but are currently limited by the sensitivity of the tracer CO emission to optical depth effects. Obviously much more work needs to be done to constrain star-formation theories, which are so critical to our understanding of galaxies.

### Active galactic nuclei and black holes

What is the role of active galactic nuclei (AGN) in galaxy formation and evolution? The relationship between the mass of the supermassive black holes and the mass of their surrounding spheroids is one of the most remarkable relationships in astrophysics and suggests an underlying connection between galaxies and black holes. However, is this relationship universal across all spheroid masses? The answer appears to be yes at the low-mass end of spheroid mass (talks by Roberto Saglia and Davor Krajnovic) but perhaps not at the high-mass end. Where does this relationship come from? It could be due to a self-limiting cycle of black hole growth, followed by kinetic energy from the AGN quenching both star formation and further accretion by the black hole. Active galaxies appear to have a greater incidence of streaming motions in the large-scale gas distributions (Gaelle Dumas), which then might reach the nucleus through torques on the gas (unlikely), but perhaps viscosity is a more likely mechanism, and finally this may lead to both star formation and accretion onto the black hole (Eva Schinnerer). But what comes first, the supermassive black hole or the galaxy? Dynamical masses derived from CO observations of very high-redshift powerful AGN (Figure 4) suggest that supermassive black holes become very massive before their galaxies have grown substantially. In fact, instead of being about 0.1 % of the total bulge mass, the black hole mass at $z \sim 6$ it is more like 3 % (Fabian Walter)! When they do get fuelled, the AGN, or at least the radio-loud AGN, can drive vigorous outflows of the type necessary to suppress further star formation and black hole growth (Nicole Nesvadba). But it also appears as if the halos of radio galaxies might have significant amounts of H I, up to $10^{11-12}$ solar masses, as probed by resonantly scattered Ly-alpha radiation (Joshua Adams).

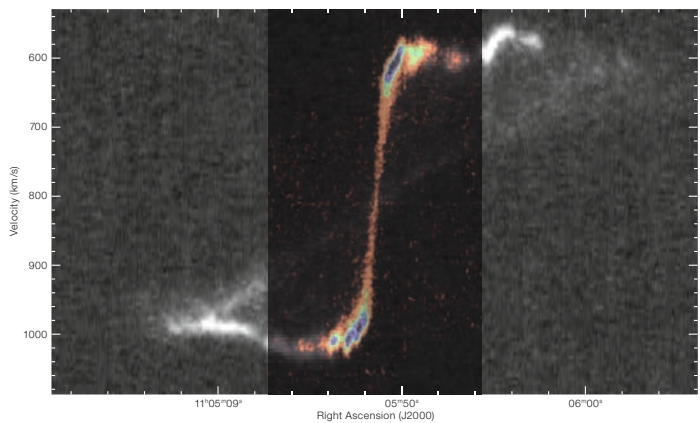

Figure 3. Major axis position-velocity diagram of NGC 3521. Greyscales are neutral hydrogen 21-cm emission from the THINGS project (Walter et al., 2008). The colours represent HERA CO emission (HERACLES, PI: Adam Leroy). Picture credit: Erwin de Blok.

What processes drive the growth of mass in galaxies? What are the relative roles of gas accretion, from the dark matter halo and surroundings, versus merging with other galaxies as the driving force for the star-formation history of the Universe? This debate was joined in the meeting from several different directions. Direct observations of gas accretion in nearby galaxies are scant. The amount of H I seen in the halos of galaxies is relatively insignificant (talks by Filippo Fraternali, George Heald and Tobias Westmeier) and its origin is unclear (George Heald). Perhaps these are not the correct type of observations, and this accreting gas is in another phase yet to be probed. The role and nature of mergers in the local Universe is of course not disputed. The most actively star-forming galaxies in the local Universe are gas-rich mergers and provide an important testing ground for our theories of star formation in active environments (presentations by Christine Wilson and Susanne Aalto), even if we do not understand completely how they evolve (John Hibbard). In such environments, with their high optical depths, it is important to probe the gas in a number of molecular species, thus providing information about the physical conditions within the gas (talks by Susanne Aalto and Masatoshi Imanishi). Mergers may also play an important role in our understanding of the environments of galax-

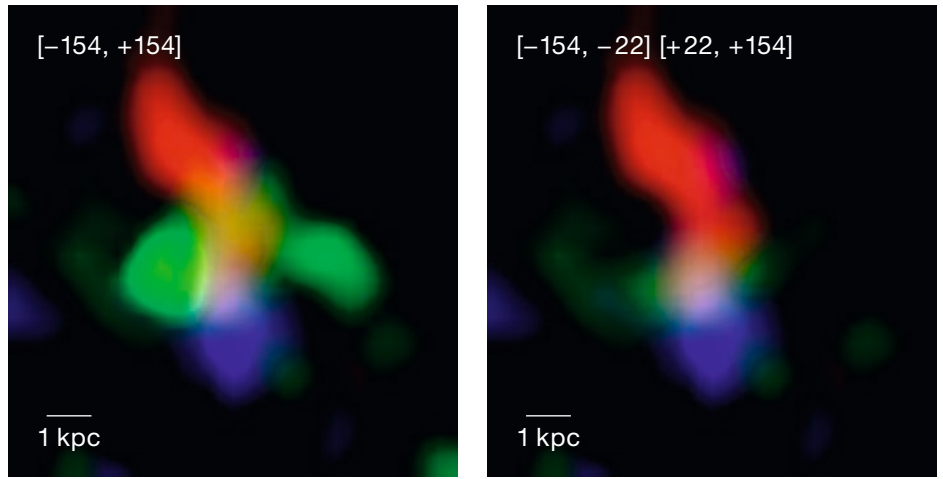

Figure 4. CO velocity field of the $z = 4.4$ interacting galaxy BRI 1335 (Riechers et al., 2008), with the velocity of the gas colour coded. The left panel shows the whole velocity range (indicated in the top), while in the right panel the contribution of gas around the systemic velocity of the galaxy is excluded, emphasising the outflows.

ies. Tidal dwarfs, formed out of the tidal arms that are generated during the merger process, helping us to understand the properties of star formation in differing environments, can also provide important information about the nature of dark-matter halos (Pierre-Alain Duc). Three-dimensional observations of high-redshift galaxies ($z \sim 2$) in the rest-frame optical and mm, on the other hand, emphasised the role of gas accretion. Some participants suggested that these observations are apparently not consistent with a significant role of mergers (Kristen Shapiro), but rather with a simple settling of discs over many dynamical times (Natascha Förster Schreiber, see Figure 5). The star-formation rates in high-$z$ galaxies appear to be consistent with gas accretion from hierarchical merging models, as well as the star-formation properties like local discs (talks by Nicolas Bouché, Helmut

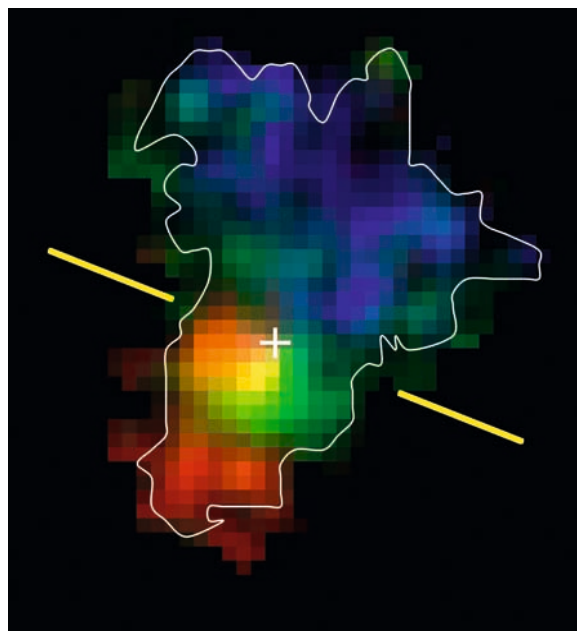

Figure 5. H-alpha velocity field of the $z = 2.38$ galaxy BzK-15504 obtained with SINONFI+AO (from Genzel et al., 2006).

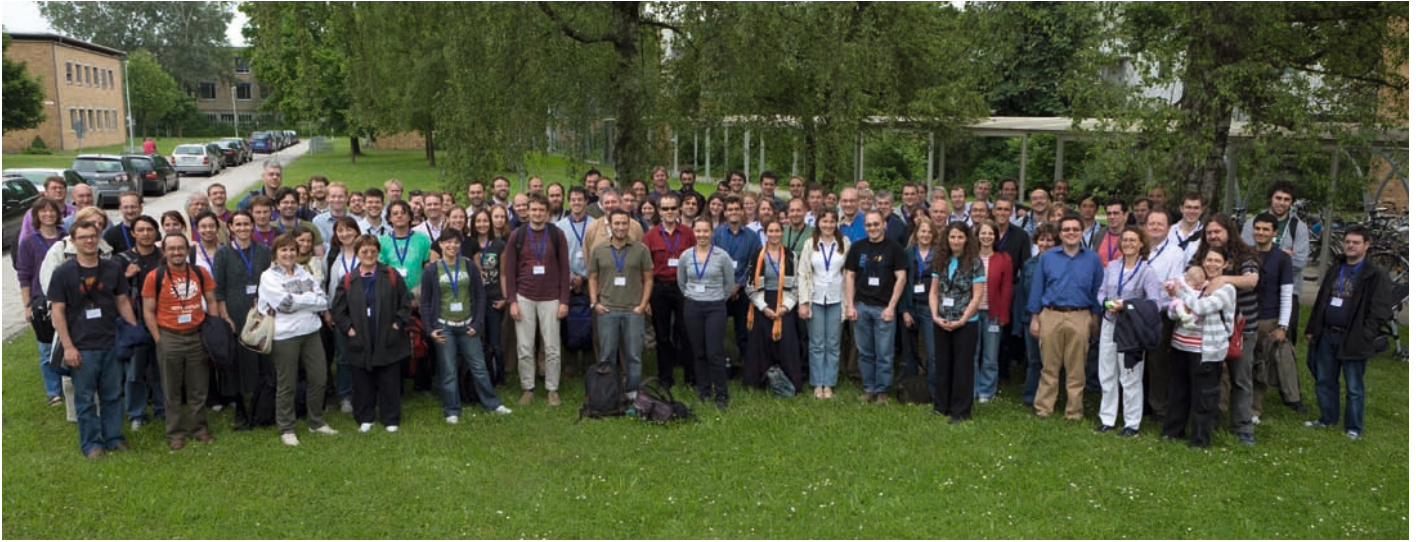

Figure 6. The conference participants group photograph taken just in front of the Max-Planck-Institut für Plasmaphysik lecture theatre.

Dannerbauer and Mark Swinbank). However, it also appears that there is a great deal of diversity in the observed molecular properties of high-redshift populations (Pierre Cox).

Of general concern with high-redshift galaxy studies is the low spatial resolution. At $z = 2$, the scale is about 8 kpc per arcsecond; thus with seeing-limited, or even adaptive-optics, observations the resolution is no better, and often much worse, than 1 kpc. Strongly lensed galaxies, however, offer the opportunity to obtain physical resolutions of ~ 100 pc and investigate the fine-scale relationships between the optical emission-line gas and the molecular gas (as exemplified in talks by Mark Swinbank and Andrew Bunker). As such, this is a powerful technique for studying the phenomenology of distant galaxies, such as their ability to drive winds, and investigating whether or not their star formation is similar to that in the local Universe. In the mm regime, ALMA will have a significant impact (talk by Robert Laing), as is already illustrated by the new extended baselines of the IRAM Plateau de Bure interferometer (Pierre Cox).

A very important component of this meeting, and one that perhaps makes it unique for a meeting of this kind, was the various talks on data reduction and visualisation techniques. Of course, mm- and cm-wave astronomers have been using 3D visualisation techniques for decades (Thijs van der Hulst) but these are still relatively new for optical and near-infrared astronomy (Giovanni Cresci). Particularly interesting are the techniques being used in medical imaging and diagnosis (described by Neb Duric). While often in a different regime (higher resolution and signal-to-noise), medicine is producing a number of powerful techniques to look for subtle relationships in three-dimensional (and four-dimensional!) data. The vast explosion in data rates in astronomy should also not be overlooked. How are we going to handle this flood of data? Visualisation should make more use of the computing power of modern Graphical Processing Units, developed for the computer game industry (Chris Fluke). Of course the *raison d'être* of the Virtual Observatory is to make this vast quantity of data, with all its complexity, available to the community (Igor Chilingarian).

In summary, it was clear from the myriad of physical processes, which must be understood in order to understand galaxies and star formation, that we have our work cut out for us. The amount of detail that the current generation of 3D facilities is revealing in galaxies is quickly advancing our knowledge. The next generation of observing facilities (e.g. ALMA and the second-generation VLT instruments) being planned or developed will only add to this happy state of affairs. However, what was also clear from the discussions during the meeting is that we need to develop our theoretical understanding and modelling techniques to be able to truly take advantage of our new observational abilities. While overall the meeting was optimistic about the future of research into gas and stars in galaxies, it was also obvious that we have a lot more to learn!